\documentclass[9pt,twocolumn,twoside]{osajnl}

\journal{ol} 

\setboolean{shortarticle}{false}

\ifthenelse{\boolean{shortarticle}}{\colorlet{color2}{color2b}}{\colorlet{color2}{color2}} 

\newcommand{\ka}{\hbox{\ae}}
\title{Generation, amplification, frequency conversion and reversal of propagation of THz photons in nonlinear hyperbolic metamaterial}
\author[1,*]{Alexander K. Popov}
\author[2,3]{Sergey A. Myslivets}
\affil[1]{Birck Nanotechnology Center, Purdue University,
West Lafayette, IN 47907,~USA}
\affil[2]{Kirensky Institute of Physics, Federal Research Center KSC SB RAS, 50, Akademgorodok, Krasnoyarsk 660036, Russia,}
\affil[3]{Siberian Federal University, 79 Svobodny  Av., Krasnoyarsk, 660041, Russia}
\affil[*]{Corresponding author: popov@purdue.edu}
\dates{Compiled \today}
\ociscodes{(190.4223)   Nonlinear wave mixing; (190.4410)   Nonlinear optics, parametric processes; (190.5530)   Pulse propagation and temporal solitons; (160.3918)   Metamaterials.}
\doi{\url{http://dx.doi.org/10.1364/ol.XX.XXXXXX}}

\setboolean{displaycopyright}{true}

\begin{abstract}%
The metamaterial is proposed which supports a mixture of three or more normal and backward electromagnetic modes with equal co-directed phase velocities and mutually contra-directed energy fluxes. This enables extraordinary three-wave mixing, greatly enhanced optical parametric amplification and frequency-changing generation of entangled photons in the reflection direction. Proof-of-principle numerical simulation of such processes is presented based on the particular example of the wave-guided THz waves contra-propagating in the metamaterial made of carbon nanotubes.
\end{abstract}
\setboolean{displaycopyright}{true}
\begin{document}
\maketitle
Backward electromagnetic waves (BEMWs) are the extraordinary waves with contra-directed phase velocity and energy flux. They do not exist in the naturally occurring materials. However, with the advent of nanotechnology, it became possible to deliberately engineer the nanostructures which produce such a significant delay of the electromagnetic response that it effectively gives rise to BEMWs \cite{Cai:OptMM.2010}. Optical BEMWs enable numerous breakthroughs in photonics, such as super-resolution and cloaking. Besides applications in the linear optics, extraordinary coherent, i.e. phase dependent, nonlinear optical (NLO) propagation processes have been predicted which hold the promise of significant advances in the harvesting of light. Among them are the processes of generation, amplification, modulation, frequency-conversion, and the propagation direction reversal of the light waves \cite{Popov:APB.84.131,Shadrivov:JOSAB.23.529,Lapine:RMP.31.129}. A main  approach to produce the optical BEMWs is  grounded on the engineering of the metamaterials (MMs) with negative refractive index \cite{Cai:OptMM.2010}. In this work, we propose a different general approach to the outlined advancements  which is based on the guided EMW with variable positive to \emph{negative} spatial dispersion  \cite{Popov:APA.109.835}.
In the case of negative spatial dispersion $\partial\omega/\partial k<0$, the Poynting vector ${\mathbf{S}}={\mathbf{v}_g}U$ ($U$ is the energy density), which is directed along the group velocity ${\mathbf{v}_g}=(\mathbf{k}/k)(\partial\omega/\partial k$), and the phase velocity, which is directed along the wave vector ${\mathbf{k}}$, appear contra-directed. The negative spatial dispersion can be realized in hyperbolic MMs \cite{BelKivsh:NPh.7.2013}. However, the key point of our work is as follows.
As shown in \cite{Popov:APB.84.131}, a huge enhancement in the coherent NLO coupling, such as in three-wave mixing, and in the related optical parametric amplification (OPA) and frequency conversion occurs only if the energy fluxes of the coupled waves are \emph{contra-directed}, whereas their phase velocities are \emph{co-directed} and \emph{equal}. This imposes a set of strict requirements to the MMs  being engineered. In this paper, we propose a general approach to engineering such MMs and present a proof-of-principle model of such a MM which operate in the THz and near-IR frequency ranges. Extraordinary  NLO processes in the given MM which give rise to the OPA and to the frequency and propagation direction changes are investigated and the outcomes are demonstrated through numerical simulations. The applications include ultra-miniature amplifiers, frequency changing reflectors and remotely actuated sensors in the THz. It is shown that the reported results can be extended to other constituent materials and frequency ranges.
\begin{figure}[ht]
\begin{center}
\includegraphics[width=.48\columnwidth]{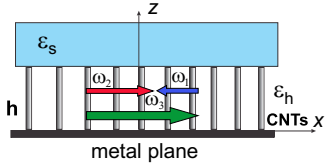}
\caption{Metamaterial composed of standing CNTs.} \label{geom}
\end{center}
\end{figure}

Consider a MM slab made of carbon nanotubes (CNTs) of height $h$ standing on the perfectly conducting metal surface and embedded in the host dielectric with the relative permittivity $\epsilon_h$. On the top, it is bounded  by the dielectric with  the relative permittivity $\epsilon_s$ (Fig.~\ref{geom}). The CNTs form a two-dimensional periodic structure in the $xy$-plane with a square lattice constant $d$. Basically, a shape of the lattice can be modified. Carbon nanotubes possess strongly anisotropic metallic property along the axis $z$ which determines  the negative relative  permittivity $\epsilon_{zz}$ \cite{Lindell:MOPTL.31.129,Nefedov:PhNano.9.374,Nefedov:PRB.84.113410}
~\begin{equation}\label{eq:p2}
 {\epsilon_{zz}}=1-{k_p^2}/({k^2+i\xi k}), \, k_p^2={\mu_0}/(d^2L_{0}).
\end{equation}
Here, $k=\omega/c<k_p$, $k_p$ is the effective plasma wavenumber, $L_{0}$ is the effective inductance of the CNTs per unit length, the parameter $\xi=1/c\tau$ is responsible for losses, $\tau$ is the electron relaxation time.  Radius $r$ of the CNTs  and the lattice constant $d$ are taken to be $r=0.82$\,nm and $d=15$\,nm. Then the indicated parameters are estimated as   $L_0=3.7\times10^{-3}$ H/m,  $k_p^2= 1.51\times 10^{12}$\,m$^{-2}$, $\omega_p/2\pi$ = 58.7~THz, $\tau$=3~ps \cite{Nefedov:PhNano.9.374,Nefedov:PRB.84.113410}. Thus the MM can be thought of as the medium with hyperbolic dispersion for the frequencies below $\omega_p$, because then $\epsilon_{zz}<0$, whereas $\epsilon_{\perp}=\epsilon_{xx} =\epsilon_{yy}=\epsilon_h>0$. As shown below, the fact that the CNTs have a limited height plays a crucial role in achieving the stated goal.
{The MM depicted in Fig.~\ref{geom}} can be also thought of as a \emph{tampered waveguide} formed by the metal plate and by the upper dielectric layer. Space-time dependence of the fields propagating along the waveguide can be approximated as $\exp{[i(k_xx-\omega t)]}$.
Maxwell equations for the allowed modes split into two  sub-systems: for the TM-modes with $H_x=0,\,H_z=0,\,E_y=0$, and the TE-modes with $E_x=0,\,E_z=0,\,H_y=0$. Equations  for the tangent components of the TM and TE modes respectively are:
\begin{equation}\label{eq:tm}\begin{array}{l}
{d}E_x(z)/{dz}=i\eta \left[k-({k_x^2}/k\epsilon_{zz})\right]H_y, \\
{d}H_y(z)/{dz}=i(k/{\eta})\, E_z=-({k_x}/{k}\epsilon_{zz})\eta H_y.
\end{array}
\end{equation}
\begin{equation}\label{eq:te}\begin{array}{l}
{d}E_y(z)/{dz}=-i\eta kH_x,\, H_z=({k_x}/{k\eta})E_y,\\
{d}H_x(z)/{dz}=({i}/{\eta})\left[({k_x^2}/{k})-k\epsilon_{\perp}\right]E_y.
\end{array}
\end{equation}
Here, $\mu=1,\eta=(\mu_0/\epsilon_0)^{1/2}$. Since $\epsilon_{zz}<0$ and $\epsilon_{\perp}>0$, only TM-modes exhibit the hyperbolic dispersion and are of interest.

The system (\ref{eq:tm}) can be presented in a matrix form:
\begin{equation}\label{eq:a4}
{d{\rm X}(z)}/{dz}=i{\rm [A]}{\rm X}(z)\end{equation}
with the column-vector ${\rm X}=(E_x,H_y)_{\rm col}$ and elements of the matrix [A] defined by Eq.~(\ref{eq:tm}).
The solution to the Cauchy problem for this matrix equation is
\begin{equation}\label{eq:a5}
{\rm X}(z)=\exp\{i{\rm [A]}z\}{\rm X}(0).
\end{equation}
Here, [M]$(z)=\exp\{i{\rm [A]}z\}$ is the transfer matrix or the propagator which connects the tangent field components in planes $z=0$ and $z$. The propagator can be computed using spectral representation of the matrix function. In our case, the matrix exponent is expressed as \cite{Lank:1969}:
\begin{equation}\label{eq:a6}
\exp\{i{\rm [A]}z\}=\sum_{l=1}^s \exp\{i\lambda_lz\}\prod_{k\neq l}
\frac{{\rm [A]}-\lambda_k{\rm [I]}}{\lambda_l-\lambda_k},\end{equation}
where $s$ is the number of different eigenvalues $\lambda_k$ of the matrix [A], and [I] is the unit matrix.
In our case, $s=2$ and
\begin{equation}\label{eq:a7}
\lambda_{1,2}^2=k_z^2=\epsilon_{\perp}\left[k^2-(k_x^2/\epsilon_{zz})\right].\end{equation}
Using Eq.~(\ref{eq:a6}) one obtains an expression for matrix [M]:
\begin{equation}\label{eq:a8}\begin{array}{l}
M_{11}(z)=\cos(k_zz),\,
M_{12}(z)=iZ\sin(k_zz),\\
M_{21}(z)={i}\sin(k_zz)/Z,\,
M_{22}(z)=\cos(k_zz),
\end{array} \end{equation}
 where $Z=E_x/H_y=\eta{k_z}/{k\epsilon_{\perp}}$ is the transverse wave impedance for the TM-waves.
The tangent field components can be recalculated from the ground plane $z=0$ to the upper plane $z=h$ using the transfer matrix while taking into account the perfect electrical conducting conditions at $z=0$:
\begin{equation}\label{eq:a9}
E_x(h)=M_{12}H_y(0),\,
H_y(h)=M_{22}H_y(0).\end{equation}
Since the area at $z>h$ is a semi-infinite material with the relative permittivity $\epsilon_s$, the tangent electric and magnetic field components at $z=h$ are connected via the transverse impedance
\begin{equation}\label{eq:a10}
Z_s=E_x(h)/H_y(h)=\eta{k_z}/{k\epsilon_s}=i\left[\eta{\sqrt{k_x^2-k^2\epsilon_s}}\right]/{k\epsilon_s}.
\end{equation}
Here, Eq.~(\ref{eq:a7}) is used and the waves are supposed to be exponentially decaying from the surface of the MM, i.e. $k_x^2>k^2\epsilon_s$.
With aid of Eqs.~(\ref{eq:a8}) - (\ref{eq:a10}), one obtains  a dispersion equation for the TM-modes propagating along the $x$-axis \cite{PopovCNT:arXiv1602.02497}:
\begin{equation}\label{eq:dis}
{k_z\tan{(k_zh)}}=({\epsilon_{h}}/{\epsilon_s}){\sqrt{k_x^2-k^2\epsilon_s}},\end{equation}
where $k_z=\sqrt{\epsilon_h[k^2-(k_x^2/\epsilon_{zz})]}$, $k_x^2>k^2$. Dispersion Eq.~(\ref{eq:dis}) reminds the one predicted in   \cite{AleksNarim:OE.14.2006} for a layered MM. It is seen that the open-end structure imposes a significant difference on the waves dispersion as compared with an unbounded uniaxial crystal \cite{Landau:ECM.en}:
\begin{equation}\label{eq:y1}
k_x^2=\epsilon_{zz}(k^2-k_z^2).\end{equation}
It also contrasts with the case of the MM made of CNTs standing between two perfectly conducting planes:
\begin{equation}\label{eq:w1}
 k_{x}^2=\epsilon_{zz}\left[k^2-(m\pi/2h)^2\right].\end{equation}
Here, $k_z=m\pi/2h$, $h$ is the height of the waveguide, and $m$ is a positive integer determining the number of field variations along the CNTs. It is seen that  all allowed discrete EM modes corresponding to different values of $m$  at $\epsilon_{zz}<0$ and $k<m\pi/2h$ appear BEMWs,   since $dk_x^2/dk^2<0$. On the contrary, Eq.~(\ref{eq:dis}) predicts the mode with a sign-changing dispersion which is of a paramount importance for enabling the coexistence of the ordinary and BEMWs. Note, the dispersion given by Eq.~(\ref{eq:dis}) is \emph{strongly dependent} on the CNTs height $h$.
\begin{table*}
  \centering
  \textbf{Table 1}\\[1ex]
\begin{tabular}{|c|c|c|c|c|c|c|c|c|}
  \hline
  Mode &  $f$, THz & $k$, $10^5$\,m$^{-1}$ & $|k_x^{''}/k|,\,10^{-3}$ & $n_g$ & $\alpha=2k_x^{''}$, $10^{-2}$$\mu$m$^{-1}$ & $L_a$, $\mu$m & $\lambda_{\rm vac}$, $\mu$m  & $\lambda_{\rm med}$, $\mu$m  \\
  \hline
  1    &  10.43    & 2.186                     & 2.26                     & $1.45$  & 0.988    & 2331        & 28.74               & 19.86 \\
  \hline
  2    &  34.95    & 7.325                     & 17.8                     & $-7.71$ & 2.61   & 88.2        & 8.58                & 5.93 \\
  \hline
  3    &  45.38   & 9.511                     & 15.46                    & $-6.29$ & 2.94    & 78.3        & 6.61                & 4.57 \\
  \hline
\end{tabular}
\end{table*}

A spectrum of the allowed modes and their attenuation can be found by numerically solving Eq.~(\ref{eq:dis}) and calculating  the complex propagation constant $k_x=k_x'+ik_x''$.
\begin{figure}[htb]
\begin{center}
\includegraphics[width=.98\columnwidth]{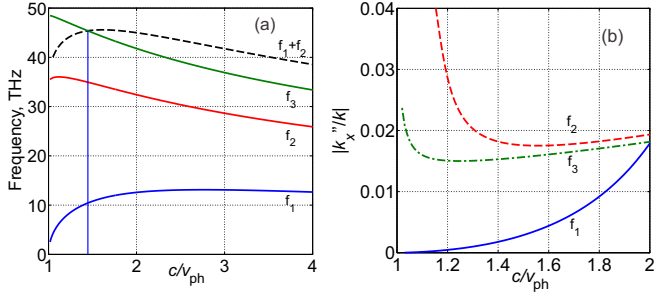}
\caption{\label{fig2}  (a) Dispersion of three lowest modes at $h=3.5\, \mu$m. The dashed line represents the sum of two lowest modes. (b) Normalized attenuation constant $k_x''/k$
: the blue (solid) line corresponds to $f_1$, the red one (dashed) -- to $f_2$, the green one ({dash-}dotted) -- to $f_3$. }
\end{center}
\end{figure}
\begin{figure}[htb]
\begin{center}
\includegraphics[width=.46\columnwidth]{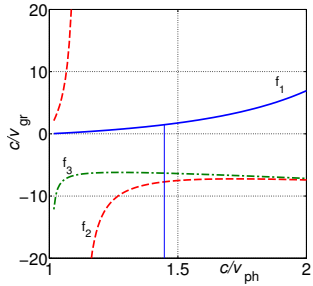}
\caption{\label{fig3} Group velocity indices $c/v_{gr}$ for the respective modes.}
\end{center}
\end{figure}
Figure~\ref{fig2}(a) presents a spectrum of the lowest modes and their dispersion $\omega(k_x')$ calculated for $h=3.5\, \mu$m. For the sake of simplicity, we assume further that $\epsilon_h=\epsilon_s=1$. Reduced wavevector $k_x'/k=c/v_{ph}$ represents phase velocity $v_{ph}$. It  proves the possibility of the phase-matching for the three-wave mixing frequency down conversion process $\omega_3-\omega_2=\omega_1$ and OPA at $\omega_2$. The simulations also prove the possibility of the phase matching for \emph{different} sets of frequencies by adjusting the nanotube lengths $h$. For convenience, a sum of the mode frequencies $\omega_1+\omega_2$ is shown by the dashed line. However, only its crossing with the third mode satisfies  frequency mixing. Phase matching  occurs at $k_x^{'}/k=1.447$ (marked with the vertical line). Figure~\ref{fig2}(b) depicts attenuation of the respective modes. As seen, the attenuations may differ greatly despite the fact that electron relaxation rate is the same. The difference is due to the nanowaveguide propagation regime.
Figure~\ref{fig3} depicts group velocity indices for the respective modes. Split for the second mode indicates slow-light regime $v_{gr}\rightarrow 0$. It is seen that in the vicinity of phase matching group velocities at $f_2$ and $f_3$ are directed \emph{against} the phase velocity with the values significantly less than the speed of light. Alternatively, group velocity for the first mode is of the opposite sign.

At the pump pulse duration $\Delta\tau=10$~ps, the pulse spectrum bandwidth is $\Delta f\approx 1/\Delta\tau=0.1$~THz, i.e., $\Delta f/f \propto10^{-2}\div10^{-3}$. Hence, the pump can be treated as quasi-monochromatic. Alternatively, for $\Delta\tau=10$~fs, $\Delta f/f \propto10$, the spectrum covers all modes.  The data used for the  simulations described below is summarized in Table~1. The values  $L_a$ indicate the metaslab length corresponding to attenuation $I_i/I_{i0}=\exp(-\alpha_i L_{ai})=0.1$, i.e., $\alpha_i L_{ai}=2.3$  for the respective frequencies. A thickness $L=L_{a3}=78.3$\,$\mu$m was chosen to satisfy   $\alpha_3L_{a3}\approx2.3$ for the mode with highest attenuation. For the same slab thickness, $\alpha_{a2}L=2.04$, and $\alpha_{a1}L=0.077$. Therefore, attenuation at the lowest frequency appears significantly less than for the two others which are comparable.
 Pump pulse length for $\Delta\tau=10$\,ps is $l=\Delta\tau v_{3gr}=\Delta\tau (c/n_{3gr})=477\ \mu$m, i.e., about 6 times greater than $L_{a3}$. Hence, in this case, a quasi-stationary process establishes through the major part of the pulse. Transient processes at the forefront and at the tail of the pulse can be neglected.

Normalized amplitudes of the waves  are given by the following equations which account for the fact that propagation direction of the wave at $f_1$ must be opposite to others in order to satisfy the phase matching:
\begin{eqnarray}
({\partial a_1}/{\partial \xi})-({v_3}/{v_1})({\partial a_1}/{\partial \tau})=
     -igla_3a_2^*+({\widetilde{\alpha}_1}/{2d})a_1,\qquad \label{eqa1}\\
 ({\partial a_2}/{\partial \xi})+ ({v_3}/{v_2})({\partial a_2}/{\partial \tau})=
  igla_3a_1^*-({\widetilde{\alpha}_2}/{2d})a_2,\qquad \label{eqa2}\\
   ({\partial a_3}/{\partial \xi})+({\partial a_3}/{\partial \tau})=
    ig^*la_1a_2-({\widetilde{\alpha}_3}/{2d})a_3.\qquad \label{eqa3}
\end{eqnarray}
Here, $\xi=x/l$, $l= v_{3}\Delta\tau$, $\tau=t/\Delta\tau$, $d=L/l$,  $\widetilde \alpha_{i}=a_{j}L$,  $v_i>0$ and $\alpha_{i}$ are the modules of  group velocities and attenuation indices at the corresponding frequencies, $g=\ka E_{30}$, $E_{i0}=E_i(x=0)$,  $\ka=4\pi\sqrt{k_1k_2} \chi^{(2)}_{\rm eff}$,  $\chi^{(2)}_{\rm eff}$ is  effective nonlinear susceptibility, $a_{i}=\sqrt{|\epsilon_i\epsilon_3|/k_ik_3}(E_i/E_{30})$.  Quantities  $|a_i|^2$ are proportional to the time dependent photon fluxes. The input pulse shape at $f_3$ was chosen to be close to  rectangular form
\begin{equation}
F(\tau)=0.5\left(\tanh\frac{\tau_0+1-\tau}{\delta\tau}-\tanh\frac{\tau_0-\tau}{\delta\tau}\right),
\end{equation}
where $\tau=t/\Delta\tau$, $\delta\tau$ is the duration of the pulse front and tail, and $\tau_0$ is the shift of the front relative to $t=0$. The values $\delta\tau=0.01$ and $\tau_0=0.5$ were taken for the simulations. The input signal at $f_2$ was supposed to be a continuous wave with $a_{20}~=~10^{-5}a_{30}$.
\begin{figure}[htb]
\begin{center}
\includegraphics[width=.98\columnwidth]{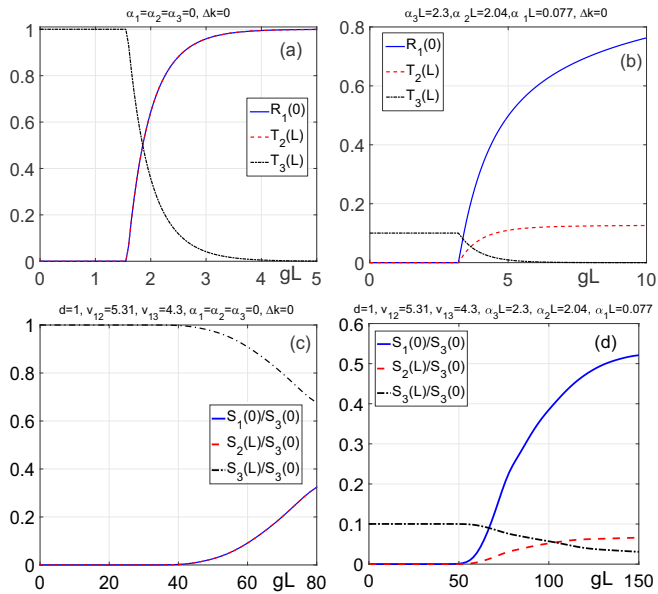}
\caption{\label{fig4} Optical parametric amplification and frequency-shifting NLO reflectivity vs. intensity of the pump.}
\end{center}
\end{figure}
Figure~\ref{fig4} demonstrates the dependence of the output amplified signal at $f_2$ and of the idler at  $f_1$ generated in the opposite direction on the intensity of the pump as well as the effects of attenuation and the pump pulse duration on the outputs. Panels (a) and (b) depict the cases of a long pump pulse (quasi-CW regime).  Here, $T_{2,3}(x=L)= |a_{2,3}(L)/a_{30}|^2$ are the transmission factors, and $a_{30}$ is the input pump maximum. The value $T_2(L)$ represents OPA. Nonlinear optical reflectivity (NLOR) at $f_1$ is given by the value $R_{10}= |a_{1}(x=0)/a_{30}|^2$.  Panel (a), which corresponds to the attenuation-free regime, shows that a huge enhancement in the OPA and  in the NLOR occurs when the pump intensity reaches a certain threshold value. This is the effect specific to BW coupling which originates from the appearance of the intensity-resonant  distributed NLO coupling feedback  \cite{Popov:APB.84.131}. Here, photon conversion efficiency reaches 100\% at a relatively small increase of the pump above the threshold value. Attenuation significantly decreases OPA (panel (b)). Remarkably, NLOR does \emph{not} experience such a significant decrease.  It is because the reflected wave is predominantly generated near the MM entrance where the pump and the signal are not yet significantly attenuated. Besides that, the attenuation for the $f_1$ mode appears to be significantly less than the one for the two other modes.
Panels (c) and (d) depict the case of a shorter pulse $l=L$. Here, the transmitted and reflected photon fluxes per pulse are given by the values $S_i(x)/S_{30}=\int dt |a_i(x,t)|^2/\int dt|a_{30}(t)|^2$. It is seen that the dispersion and the opposite signs of group velocities cause a significant increase of the pump threshold  and a decrease of the sharpness of the enhancement in the vicinity of the threshold.
\begin{figure}[t!]
\begin{center}
\includegraphics[width=.99\columnwidth]{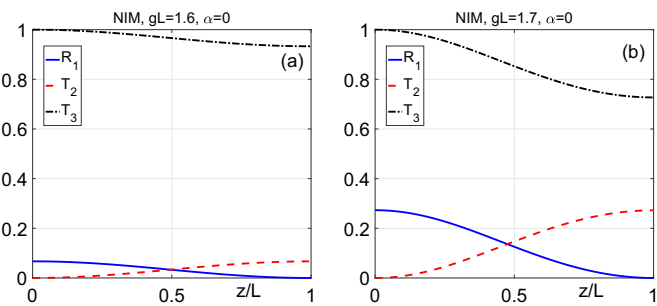}
\caption{\label{fig5} Fields distribution along the attenuation-free metaslab  in the vicinity of the resonance pump intensity.}
\end{center}
\end{figure}
\begin{figure}[t!]
\begin{center}
\includegraphics[width=.99\columnwidth]{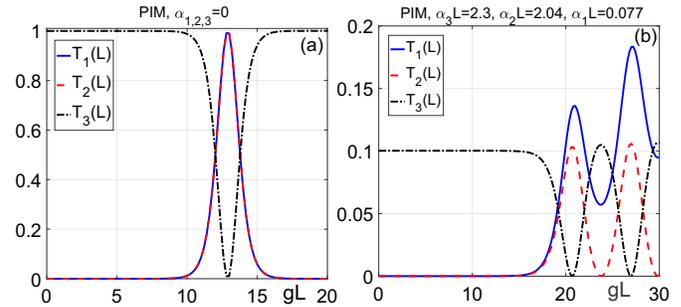}
\caption{\label{fig6} Output fields vs intensity of the pump for the ordinary, co-propagating coupling scheme. All other parameters are the same as in Figs.~\ref{fig4} (a) and (b) respectively.}
\end{center}
\end{figure}
The described dependencies are due to the unusual Manley-Rowe relationship. A \emph{difference} of the pump and of the \emph{contra-propagating} idler photon numbers is a constant along the slab,  whereas  the \emph{sum} of the pump and of the \emph{co-propagating} signal photon numbers is a constant as seen in Fig.~\ref{fig5} calculated for the ultimate case of the attenuation-free  metaslab. A width of the gap between the pump and the idler curves represents a conversion rate. The gap sharply decreases and the conversion rate increases in the vicinity of the threshold pump intensity.

Figure~\ref{fig6} is calculated for the co-propagating coupling in the ordinary material with the same other parameters as in Figs.~\ref{fig4} (a) and (b). It is seen that higher conversion efficiency at lower pump intensity is achieved in the first BEMW coupling case.

{\textbf{CONCLUSIONS:}}
The possibility  of engineering the metamaterials which enable a great enhancement of coherent nonlinear optical coupling of  electromagnetic waves is shown. This allows for the decreasing of the size of a material slab which is needed for the efficient frequency conversion and optical parametric amplification of signals. The proposed metamaterials support a set of the guided  electromagnetic modes whose dispersion $\omega(k)$  can be adjusted so that  their frequencies would satisfy the wave mixing while their propagation phase velocities become equal. The latter is the prerequisite of a fundamental importance for efficient nonlinear optical frequency mixing. The even more important point is that the proposed metamaterials enable the modes with oppositely directed energy fluxes while their phase velocities are co-directed and equal. Indeed, the latter is a key factor that ensures a great enhancement in the wave coupling. Besides the enhanced  parametric amplification of  weak signals and  compensating their losses, such an uncommon coupling scheme enables generation of the extraordinary frequency shifted idler in the direction opposite to the signal.  Such possibility is demonstrated  through numerical simulations of the corresponding  propagation processes with the aid of a particular model of the metamaterial suitable for the THz frequency range. The outlined properties of the investigated processes hold promise for  applications in the photonic device technologies and in the remote sensing.

We thank I. S. Nefedov for helpful discussions.\\
{\textsf{\textbf{Funding}}.}
U. S. Army Research Office (W911NF-14-1-0619); Russian Foundation for Basic Research (RFBR 15-02-03959A).

\end{document}